\documentclass[aps,prb,twocolumn,superscriptaddress]{revtex4-2}

\usepackage{graphicx}
\usepackage{mathrsfs}
\usepackage{bm}
\usepackage{amsmath}
\usepackage{color}
\usepackage{dcolumn}
\usepackage{epstopdf}
\usepackage{dsfont}
\usepackage{amssymb}
\usepackage{tabularx}
\usepackage{array,ulem}
\usepackage{float}
\usepackage{colordvi}
\usepackage{verbatim}
\usepackage{hyperref}
\hypersetup{
	colorlinks=true,
	linkcolor=blue,
	filecolor=blue,
	urlcolor=blue,
	citecolor=blue,
}

\begin{document}
\title{Quantum spin Hall effect in bilayer honeycomb lattices with $C$-type antiferromagnetic order}

\author{Lizhou Liu}
\affiliation{College of Physics, Hebei Normal University, Shijiazhuang 050024, China}

\author{Cheng-Ming Miao}
\affiliation{International Center for Quantum Materials, School of Physics, Peking University, Beijing 100871, China}

\author{Qing-Feng Sun}
\affiliation{International Center for Quantum Materials, School of Physics, Peking University, Beijing 100871, China}
\affiliation{Hefei National Laboratory, Hefei 230088, China}

\author{Ying-Tao Zhang}
\email[Correspondence author:~~]{zhangyt@mail.hebtu.edu.cn}
\affiliation{College of Physics, Hebei Normal University, Shijiazhuang 050024, China}

\date{\today}

\begin{abstract}
We propose a scheme to realize time-reversal symmetry-broken quantum spin Hall insulators using bilayer honeycomb lattices, combining intrinsic spin-orbit coupling, $C$-type antiferromagnetic ordering, and staggered potentials. The $C$-type antiferromagnetic order emerges from the interplay between intralayer antiferromagnetism and interlayer ferromagnetism. The system's topological properties are characterized by the spin Chern number. We present the topological phase diagram of the bilayer honeycomb lattice, providing a detailed insight into the stability and tunability of the quantum spin Hall effect in this system. The presence of helical edge states is confirmed by the measurement of quantized longitudinal resistance values of $3/2(h/e^2)$ and $1/2(h/e^2)$ in a six-terminal Hall-bar device. Remarkably, this quantum spin Hall insulator phase is protected by interlayer parity-time ($\mathcal{PT}$) symmetry, despite the breaking of time-reversal symmetry.
\end{abstract}

\maketitle
\section{Introduction}

Research on topological phases began with the groundbreaking discovery of the integer quantum Hall effect, a seminal result in condensed matter physics that revealed the profound connection between topology and electronic properties~\cite{Haldane1988, Chiu2016, Bansil2016, Hasan2010, LiuFeng, Qi2011, Thouless1982, Xiao2010, Teo2010}. This discovery laid the foundation for understanding how topological invariants, which classify different quantum states, play a crucial role in determining the behavior of electronic systems.
Building on this, the subsequent theoretical prediction and experimental realization of the quantum spin Hall (QSH) effect marked a major milestone in the field of topological insulators~~\cite{Kane2005, Kane2005a, Bernevig2006, Bernevig2006a}. Unlike the integer quantum Hall effect, associated with charge transport, the QSH effect exhibits topologically protected helical edge states, enabled by spin-momentum locking arising from time-reversal symmetry and spin-orbit coupling.
The $Z_2$ topological invariant, directly linked to time-reversal symmetry, characterizes these phases.
In addition to the $Z_2$ index, the spin Chern number, defined as $\mathcal{C}s = \frac{1}{2}(\mathcal{C}_\uparrow - \mathcal{C}_\downarrow)$, serves as a topological invariant characterizing the QSH effect~\cite{spinchern1, spinchern2, spinchern3}.
This property enables dissipationless spin transport, making QSH insulators highly suitable for low-power magnetic memory devices.
However, magnetism disrupts the time-reversal symmetry and leads to novel quantum phenomena, including the quantum anomalous Hall effect~\cite{Yu2010, Liu2008}, higher-order topological phases~\cite{Ren2020, Liu2024, Miao2022, Miao2024, Liu2024a}, and axion insulators~\cite{Liu2020}.

On the other hand, antiferromagnetism, a distinctive magnetic order that breaks time-reversal symmetry, has garnered significant attention in spintronics, giving rise to the concept of antiferromagnetic electronics~\cite{Jungwirth2016, Smejkal2018, Baltz2018}.
The unavoidable presence of magnetism in certain materials has driven the exploration into the magnetic QSH effect.
Three-dimensional antiferromagnetic topological insulators have been extensively studied~\cite{Mong2010, Fang2013, Liu2014, Zhang2019, Li2019, Otrokov2019, Otrokov2019a} and have been experimentally realized in $\rm MnBi_2Te_4$\cite{Otrokov2019a}.
Furthermore, a nonsymmorphic-protected two-dimensional antiferromagnetic topological insulator has been proposed, exhibiting helical edge modes with spin-momentum locking~\cite{Niu2020, Mao2020}.

\begin{figure}
  \centering
  \includegraphics[width=8.5cm,angle=0]{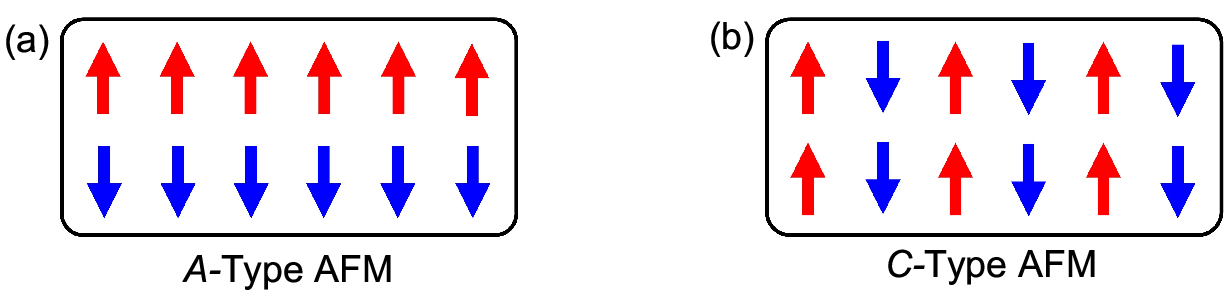}
 \caption{Two types of antiferromagnetic structures, (a) $A$-type antiferromagnetic, and (b) $C$-type antiferromagnetic. AFM denotes antiferromagnetic.
 }
  \label{fig1}
\end{figure}

In contrast, two-dimensional bilayer materials exhibit a wider variety of antiferromagnetic orders~\cite{Li2022, Jang2023, Feng2023, Gao2021, Peng2023, Tong2017, Gao2020, Mu2023, Wu2023, Sun2022, Ren2022, Peng2020, Zhao2023, Guo2020}, owing to the additional layer degrees of freedom.
Recently, the $A$-type antiferromagnetic QSH effect has garnered attention~\cite{Xue2023, Tian2024, Wu2024}.
$A$-type antiferromagnetism refers to a combination of intralayer ferromagnetism and interlayer antiferromagnetism, as illustrated in Fig.~\ref{fig1}(a).
In addition to $A$-type antiferromagnetism, $C$-type antiferromagnetism is characterized by intralayer antiferromagnetism and interlayer ferromagnetism, as illustrated in Fig.~\ref{fig1}(b).
Due to its antiferromagnetic properties within each layer, $C$-type antiferromagnets are more robust against external magnetic fields than $A$-type antiferromagnets.
However, the exploration of $C$-type antiferromagnetic QSH insulators remains largely unexplored.

In this work, we propose a scheme to realize $C$-type antiferromagnetic QSH insulators using bilayer honeycomb lattices.
As shown in Fig.~\ref{fig2}(a), the sublattices A and B of the bilayer honeycomb lattice are represented by the red and blue spheres, respectively, and the arrows indicate the direction of the out-of-plane exchange field.
The magnetic properties of the top and bottom honeycomb lattices are identical, featuring compensating antiferromagnetism in the A and B sublattices with reversed parallel alignment, which is characteristic of $C$-type antiferromagnetism\cite{Yuan2019}.
Furthermore, intrinsic spin-orbit coupling and staggered potentials in the bilayer honeycomb lattice are considered to realize a QSH insulator.
The nonzero spin Berry curvature $\Omega$ and spin Chern number $\mathcal{C}_s$ serve as topological invariants of this QSH effect.
Additionally, we present a topological phase diagram for the $C$-type antiferromagnetic bilayer honeycomb lattice.
We then calculate the energy band structure and wave function distribution of zigzag nanoribbons, demonstrating the existence of spin momentum-locked helical edge states within the bulk gap.
Finally, using a six-terminal transport device, we reconfirm the existence of these helical edge states, with quantized transport coefficients obtained from the Landauer-B$\ddot{\rm u}$ttiker formalism.

\section{System Hamiltonian}

\begin{figure}
\centering
\includegraphics[width=8.5cm,angle=0]{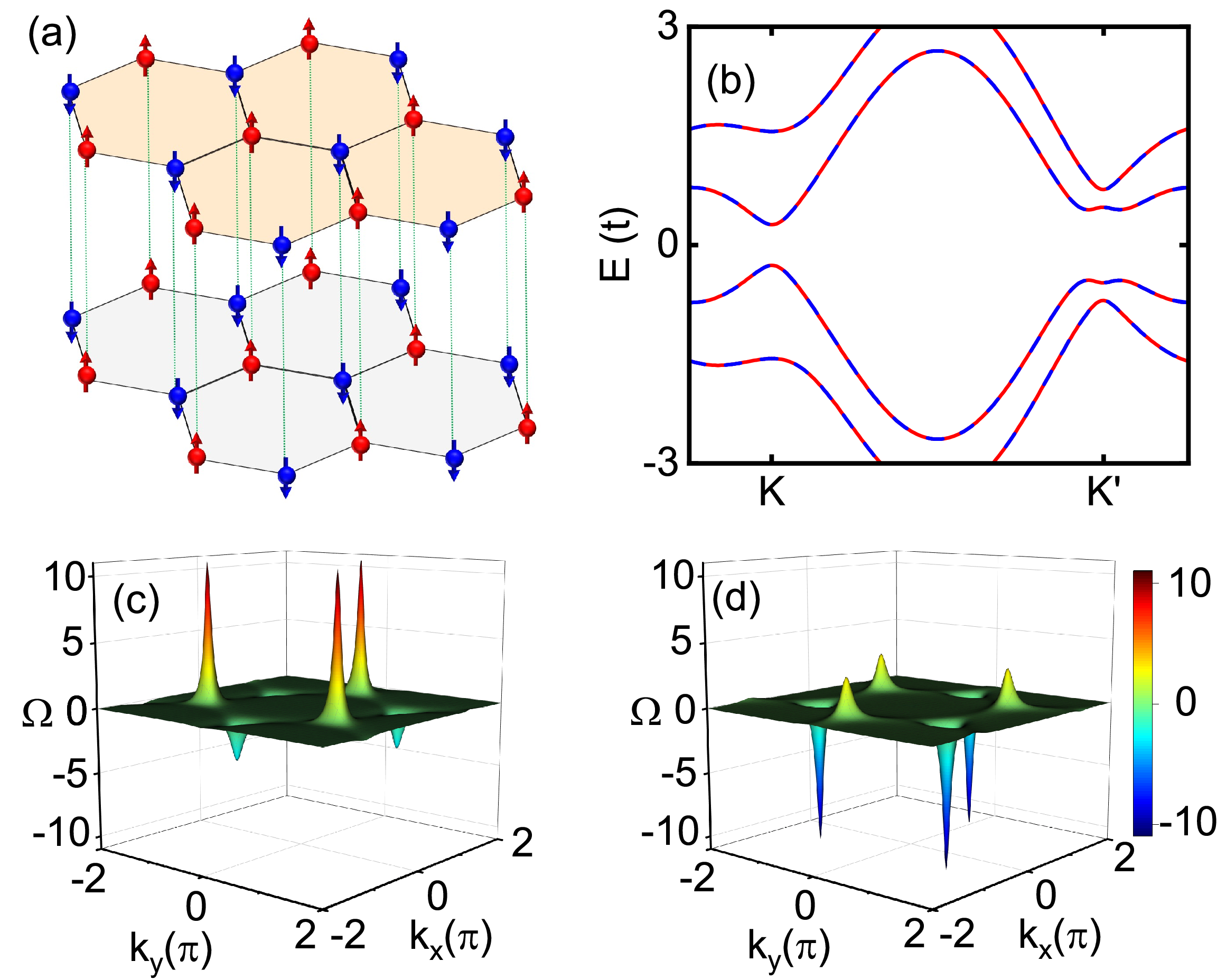}
\caption{(a) Crystal structure of the $C$-type antiferromagnetic bilayer honeycomb lattice. Red and blue arrows represent the spin directions on sublattices A and B.
(b) Bulk energy spectrum of the bilayer honeycomb lattice model.
The spin-up (red lines) and spin-down (dotted blue lines) bulk bands are gapped and degenerate.
(c) and (d) Berry curvature distribution $\Omega$ (in units of $e^2/h$) of the valence bands in the momentum space for spin-up (c) and spin-down (d).
Parameters are chosen as $t = 1$, $t_{\rm{I}}  = 0.1$, $\lambda = 0.3$, $\mu=0.4$, and $t_\perp=0.4$.}
\label{fig2}
\end{figure}

The tight-binding Hamiltonian of the bilayer honeycomb lattices
as shown in Fig.~\ref{fig2}(a) is given by:
\begin{eqnarray}
H &=&-t\sum_{\langle ij \rangle}c^\dagger_{i}c_{j} + i t_{\rm{I}} \sum_{\langle\langle ij \rangle\rangle}\nu_{ij}c^\dagger_{i}{s}_{z}c_{j}+\lambda \sum_{i}c^{\dagger}_{i} \sigma_z s_z c_{i}  \nonumber
\\
&& +\mu \sum_{i}c^{\dagger}_{i} \sigma_z \tau_z c_{i} +t_\perp \sum_{i}c^{\dagger}_{i} \tau_x c_{i},
\label{EQ1}
\end{eqnarray}
where $c^\dagger_{i}$ $(c_{i})$ is the creation (annihilation) operator for an electron at site $i$.
The $s_i, \sigma_i,$ and $\tau_i$ $(i = x, y,  z)$ denote the Pauli matrices acting on the spin, sublattice, and interlayer degrees of freedom, respectively.
The parameter $t$ corresponds to the nearest-neighbor hopping amplitude.
The $t_{\rm{I}}$ describes the intrinsic spin-orbit coupling involving next-nearest-neighbor hopping, with $\nu_{ij}=({\hat{\bm{d}}_i \times \hat{\bm{d}_j}})_z/{|\hat{\bm{d}}_i \times \hat{\bm{d}}_j|}$, where $\hat{\bm{d}}_{i,j}$ are unit vectors along the two bonds the particle traverses going from site $j$ to $i$.
The third term represents an out-of-plane $C$-type antiferromagnetic order of magnitude $\lambda$, which could break the time-reversal symmetry.
$\mu$ is staggered potentials with sublattices $A,$ and $B$ having opposite signs, and with top and bottom layers of opposite sign, which could break the inversion symmetry.
The final term is the interlayer coupling strength of bilayer honeycomblattices.
Throughout this work, the Fermi level, intrinsic spin-orbit couplings, Zeeman field, staggered potentials, and interlayer coupling are expressed in the unit of $t$.

To further analyze the model, we transform the Hamiltonian to momentum space. The momentum space Hamiltonian can be expressed as follows:
\begin{eqnarray}
H(\textbf{k}) &=&[f_x(\textbf{k}) \sigma_x + f_y(\textbf{k}) \sigma_y ] s_0 \tau_0  + f_{\rm I}(\textbf{k}) \sigma_z s_z \tau_0 + \lambda \sigma_z s_z \tau_0 \nonumber \\
& + &\mu \sigma_z s_0 \tau_z + t_\perp \sigma_0 s_0 \tau_x,
\label{EQ2}
\end{eqnarray}
where
\begin{eqnarray}
f_x(\textbf{k}) &=&t [\cos a k_y +2\cos({ a k_y/2}) \cos({\sqrt{3}a k_x/2})], \nonumber  \\
f_y(\textbf{k}) &=& t [\sin a k_y - 2 \sin({ a k_y/2}) \cos({\sqrt{3}a k_x/2})], \nonumber  \\
f_{\rm I}(\textbf{k}) &=&- 2 t_{\rm{I}}[ \sin{\sqrt{3} a k_x} - 2 \cos({3 a k_y/2}) \sin({\sqrt{3} a k_x/2})], \nonumber
\label{EQ3}
\end{eqnarray}
with $\textbf{k}= (k_x, k_y)$ being quasi momentum and $a$ being the lattice constant.
$s_0$, $\sigma_0$, and $\tau_0$ are the $2\times 2$ unit matrices for the spin, sublattice, and layer spaces, respectively.

Notably, the antiferromagnetic order breaks the time-reversal symmetry $\mathcal{T}= i s_y \mathcal{K}$ (where $\mathcal{K}$ is complex conjugation), but preserves the $\mathcal{PT}$ symmetry $(\mathcal{PT} = i \sigma_x s_y \mathcal{K})$.
In addition, the intralayer staggered potential breaks the $\mathcal{PT}$ symmetry of the monolayer honeycomb lattice.
However, the opposite staggered potentials between the layers ensure that the top and bottom layers are each other's counterparts for $\mathcal{PT}$ symmetry, and thus the bilayer system satisfies the layer $\mathcal{PT}$ symmetry $(\tau_x \mathcal{PT})$.
Therefore, the bulk band is spin degenerate for all values of $\textbf{k}$.

\section{RESULTS AND DISCUSSION}

\subsection{Topological bulk band structure}

Next, we turn to the numerical analysis. We first examine the bulk band structure of the bilayer system using the momentum-space Hamiltonian.
Figure~\ref{fig2}(b) shows the bulk band structure of the system along the $k_y=0$ profile, with the parameters set to $t = 1$, $t_{\rm{I}} = 0.1$, $\lambda = 0.3$, $\mu=0.4$, and $t_\perp=0.4$.
The bulk band gap near the Fermi energy can be seen, which indicates that the system is in an insulating state.
The K and K' valleys appear asymmetric due to the antiferromagnetic ordering of the A and B sublattices within the layer.
Furthermore, since the out-of-plane magnetism does not hybridize spins, we distinguish between spins in the bulk band structure, with red (blue dashed) lines representing the spin-up (down) sectors.
The top and bottom layers are interlayer $\mathcal{PT}$-symmetric concerning each other, leading to the spin degeneracy bulk band of the bilayer model.

To confirm the topologically nontrivial nature of the bulk gap, we calculate the quantized single spin-charge Hall conductance: $\sigma_{yx}= \mathcal{C}_\alpha \: e^2/h\: (\alpha=\uparrow, \downarrow)$, where $\mathcal{C}_\alpha$ is an integer known as the spin-up or spin-down Chern number~\cite{Thouless1982, Matthes2016}.
It can be calculated by the following
\begin{eqnarray}
\mathcal{C_\alpha}=\frac{1}{2 \pi}\sum_n \int_{BZ} d^2 \mathbf{k} \Omega_n,
\label{EQ4}
\end{eqnarray}
where $\mathbf{k}$ is the electron wave vector, and $\Omega_n$ is the momentum-space Berry curvature for the $n$th band~\cite{Qiao2018, Yao2004, Chang1996}
\begin{eqnarray}
\Omega_n(\mathbf{k})= -\sum_{m \neq n}  \frac{2{\rm  Im} \langle \psi_{n \mathbf{k}} | v_x | \psi_{m \mathbf{k}} \rangle \langle  \psi_{m \mathbf{k}} | v_y | \psi_{n \mathbf{k}} \rangle }{ (\omega_m - \omega_n)^2}.
\label{EQ5}
\end{eqnarray}
The summation is performed over all occupied bands below the bulk gap, with $\omega_n = E_n / \hbar$, and $v_{x,y}$ representing the velocity operator.
The absolute value of $\mathcal{C}_\alpha$ corresponds to the number of single spin gapless edge states in the bilayer system.
Figures~\ref{fig2}(c) and~\ref{fig2}(d) show the Berry curvature $\Omega$ distributions for the spin-up and spin-down sectors of the valence bands in momentum space, respectively.
We observe that the Berry curvature is not symmetric in the K and K' valleys due to the breaking of sublattice symmetry by the antiferromagnetic order.
In Fig.~\ref{fig2}(c), the Berry curvature of the spin-up sector reaches a positive peak at the K point, while in Fig. \ref{fig2}(d), the spin-down sector of the Berry curvature reaches a negative peak at the K' point.
By integrating the curvature, we obtain the single spin Chern numbers and find that $\mathcal{C}_\uparrow = - \mathcal{C}_\downarrow = 1$.
The spin Chern number $\mathcal{C}_s = \frac{1}{2} (\mathcal{C}_\uparrow - \mathcal{C}_\downarrow) = 1$ of the system, confirming that the system exhibits the QSH effect with helical edge states.

\subsection{Topological phase diagram}

\begin{figure}
  \centering
  \includegraphics[width=8.7cm,angle=0]{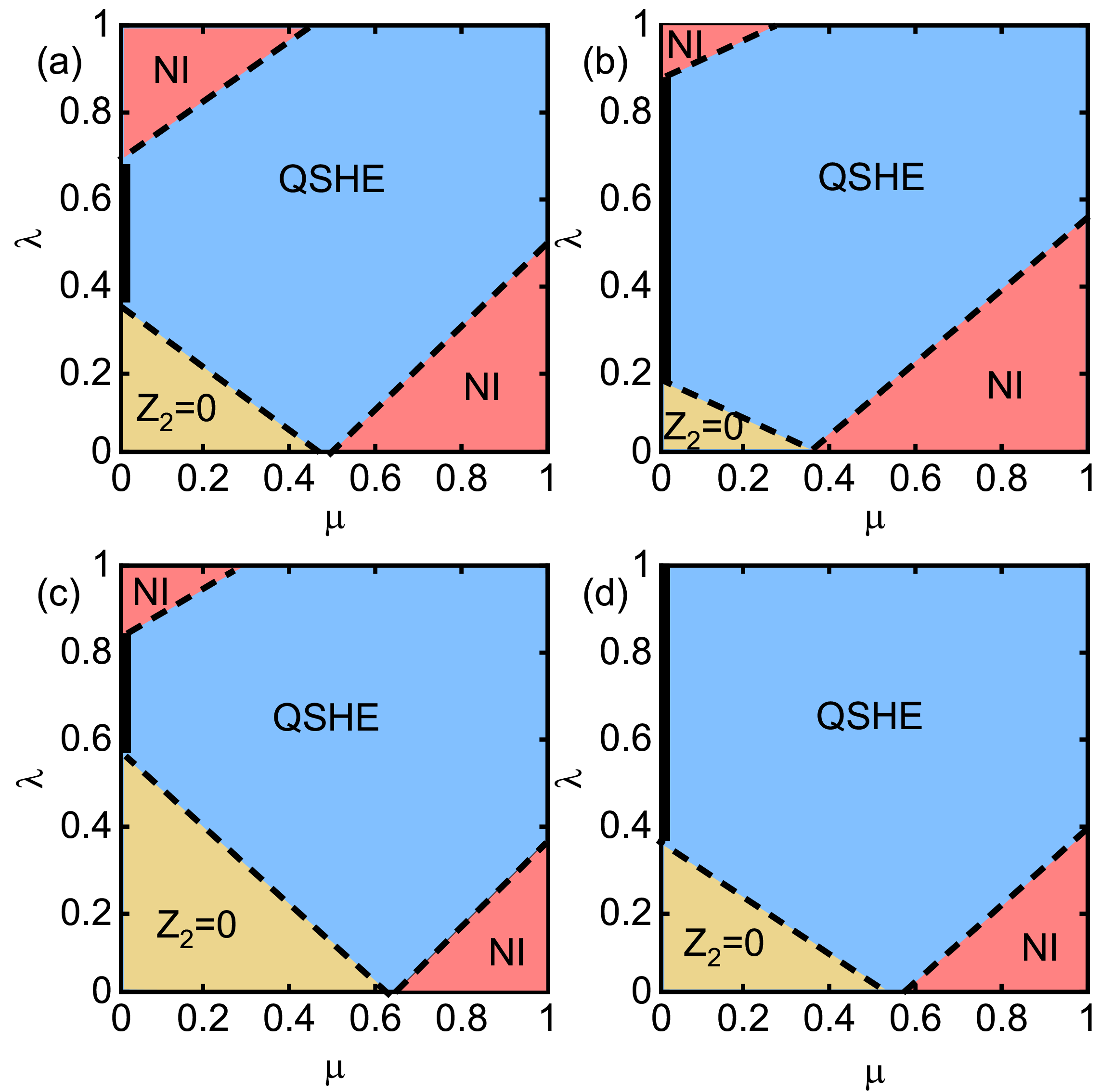}
\caption{Topological phase diagram of the $C$-type antiferromagnetic bilayer honeycomb lattice as functions of $\lambda$ and $\mu$ with $t = 1$.
The other parameters are chosen as (a): $t_{\rm{I}} = 0.1$, and $t_\perp=0.2$, (b): $t_{\rm{I}} = 0.1$, and $t_\perp=0.4$, (c): $t_{\rm{I}} = 0.13$, and $t_\perp=0.2$, and (d): $t_{\rm{I}} = 0.13$, and $t_\perp=0.4$.
The black dashed line indicates the phase boundary and the thick black solid line indicates the bulk band closure.
$Z_2 = 0$ corresponds to a weak topological insulator phase with two pairs of helical edge states, NI for normal insulator and QSHE for QSH effect.
}
\label{fig3}
\end{figure}

As described in our previous work~\cite{Liu2024b}, two honeycomb lattices models with opposite intrinsic spin-orbit couplings are coupled, resulting in a higher-order topological insulator.
However, if the intrinsic spin-orbit couplings are identical, the system exhibits a weak topological phase with two pairs of helical edge states.
In this study, we introduce $C$-type antiferromagnetism and staggered potentials in a bilayer weak topological insulator, leading to a QSH insulator with a single pair of helical edge states.

To accurately identify the phase boundaries, we plot the phase diagram as a function of the staggered potential $\mu$ and magnetization strength $\lambda$, as shown in Fig. \ref{fig3}.
In Fig. \ref{fig3}(a), four distinct regions are identified, corresponding to the QSH insulator, weak topological insulator, and normal insulator phases, with the other parameters chosen to be $t = 1$, $t_{\rm{I}} = 0.1$, and $t_\perp=0.2$.
For small values of $\lambda$ and $\mu$, the system can be viewed as two coupled honeycomb lattice models, resulting in a weak topological insulator with two pairs of helical edge states (yellow region).
As either $\lambda$ or $\mu$ reaches a critical value, the system transitions to a normal insulator (red region).
Notably, for certain simultaneous values of $\lambda$ and $\mu$, the system achieves a QSH effect with a single pair of helical edge states.
A comparison between Figs.~\ref{fig3}(a) and \ref{fig3}(b) shows that, despite an increase in interlayer coupling, the system retains the QSH phase
However, as the interlayer coupling increases, the $Z_2 = 0$ region shrinks, since interlayer coupling reduces the bulk gap, making the system more susceptible to phase transitions.
Furthermore, we investigate the effect of varying intrinsic spin-orbit couplings ($t_{\rm{I}}=0.13$) on the phase diagram, as shown in Figs.~\ref{fig3}(c,d).
By comparing Fig.~\ref{fig3}(c) and Fig.~\ref{fig3}(d) with Fig.~\ref{fig3}(a) and Fig.~\ref{fig3}(b), we observe that the region where $Z_2 = 0$ becomes larger as the intrinsic spin-orbit coupling increases. 
This suggests that a stronger intrinsic spin-orbit coupling leads to an enhanced bulk band gap, requiring greater staggered potentials $\mu$ to induce the QSH phase.
From the comparisons in Figs.~\ref{fig3}(a)-~\ref{fig3}(d), we conclude that larger intrinsic spin-orbit couplings compete with the interlayer coupling in promoting the QSH phases, especially at fixed values of $\mu$ and $\lambda$.

\subsection{Helical edge states}
\begin{figure}
  \centering
  \includegraphics[width=8.5cm,angle=0]{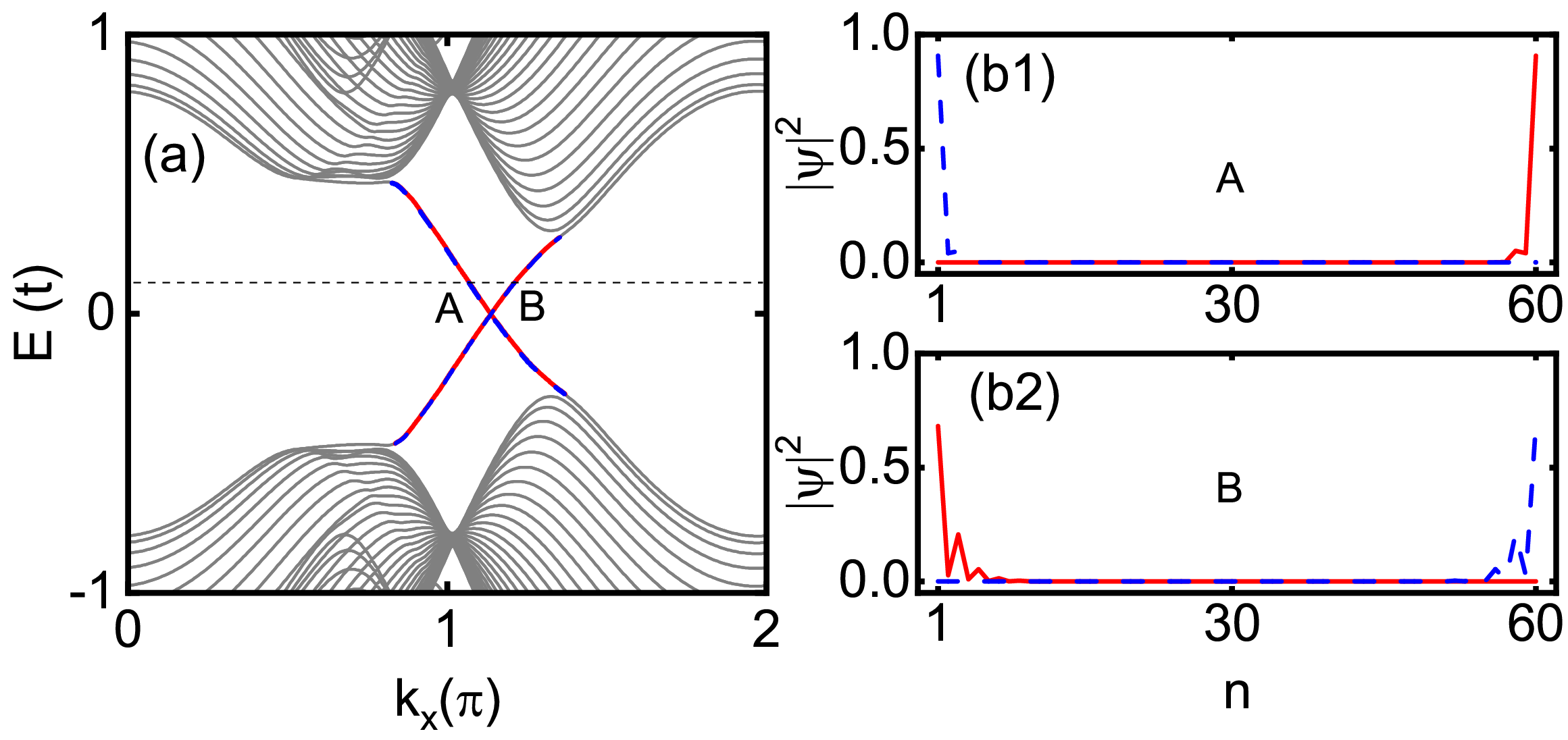}
\caption{(a) The band structure of a bilayer zigzag nanoribbon with width $N_y=60 a$.
   The Fermi level $E = 0.1$ corresponds to four different edge states for A, and B. The spin-up (red lines) and spin-down (dotted blue lines) are degenerate.
   (b1) The wave-function distributions for the two edge states at A.
   (b2) The wave-function distributions for the two edge states at B.
   The parameters are chosen as $t = 1$, $t_{\rm{I}}  = 0.1$, $\lambda = 0.3$, $\mu=0.4$, and $t_\perp=0.4$.
}
\label{fig4}
\end{figure}

We now focus on the energy band structure of the bilayer honeycomb lattice nanoribbon to reveal the helical edge states.
 In Fig.~\ref{fig4}(a), we present the band structure of a zigzag nanoribbon with a width of 60 atoms.
 The parameters are set to $t = 1$, $t_{\rm{I}} = 0.1$, $\lambda = 0.3$, $\mu = 0.4$, and $t_\perp = 0.4$.
 As shown, the K and K' valleys are asymmetric, with gapless edge states and gapped bulk states.
The edge states, which are spin-degenerate, are represented by red solid (spin-up) and blue dashed (spin-down) lines.
For a Fermi level within the gap at $E = 0.1$, two spin-degenerate edge states are observed, labeled A and B.
From $\mathbf{\nu (k)}=\frac{1}{\hbar} \frac{\partial E (k)}{\partial k}$, one can find that state A and state B propagate in the $+x$ and $-x$ directions, respectively.
Figure~\ref{fig4}(b1) shows the wave function distributions of the edge states at point A for different spins.
The spin-up (spin-down) edge state is localized at the right (left) end, as indicated by the red (blue dashed) line.
In contrast, the spin-up edge state at point B is localized at the left end, as shown in Fig.~\ref{fig4}(b2), forming a spin-up chiral edge state.
 Meanwhile, the spin-down edge state at point B is localized at the right end, forming an edge state with opposite chirality compared to the spin-up state.
This configuration is topologically equivalent to a helical edge state and distinct from a chiral edge state, even in the presence of antiferromagnetic order.
 The helical edge states obtained from the tight-binding model are consistent with a spin Chern number $\mathcal{C}_s = 1$ from the Berry curvature integral.
 Consequently, the system realizes the $C$-type antiferromagnetic QSH effect.

\begin{figure}
  \centering
  \includegraphics[width=8.6cm,angle=0]{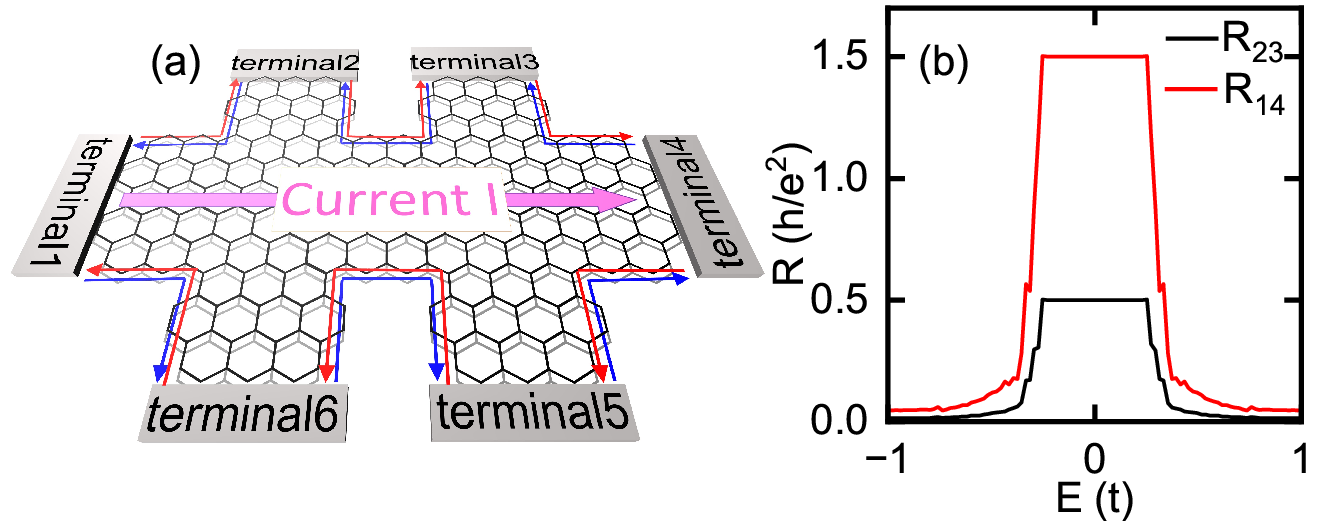}
\caption{(a) Schematic plot of a six-terminal Hall-bar device, where red and blue arrows represent spin-up and spin-down edge states, respectively. (b) The resistances $R_{23}$ (black line) and $R_{14}$ (red line) vs incident energy $E$. The widths of the terminals 1,4 (2,3,5,6) are $\frac{40a}{\sqrt{3}} (40a)$, and the size of the central scattering area is $136a*\frac{40a}{\sqrt{3}}$.
Other parameters are the same as those in Fig. \ref{fig4}.}
\label{fig5}
\end{figure}

It is well known that the helical edge states in the QSH insulator can be experimentally detected using a six-terminal Hall bar, as illustrated in Fig. \ref{fig5}(a).
Using a multiprobe Landauer-B$\ddot{\rm u}$ttiker formula \cite{Fisher1981,Metalidis2005,Sun2009}, the current $I_p$ in lead $p$ can be expressed as
\begin{eqnarray}
I_p=(e^2/h)\sum_{q(q\neq p)}T_{pq}(V_p-V_q),
\label{EQR1}
\end{eqnarray}
where $V_{p(q)}$ is the voltage in lead $p(q)$ and $T_{pq}$ is the transmission coefficient from lead $q$ to lead $p$.
In the QSH effect platform,  only the transmission coefficients $T_{p,p+1}=T_{p+1,p}=1$ will be nonzero.
Plugging this into Eq. (\ref{EQR1}) and assuming a current passing from terminal 1 to terminal 4, six linear equations about six voltages $V_p$ are obtained as:
\begin{align}
I\begin{pmatrix}
1\\0\\0\\-1\\0\\0
\end{pmatrix}=\frac{e^2}{h}\begin{pmatrix}
-2& 1&0 & 0&0 &1\\1&-2& 1&0 & 0&0 \\0&1&-2& 1&0 & 0\\0&0&1&-2& 1&0 \\0&0&0&1&-2&1 \\1&0&0&0&1&-2
\end{pmatrix}\begin{pmatrix}
V_1\\V_2\\V_3\\V_4\\V_5\\V_6
\end{pmatrix}.
\end{align}
Assuming that terminal 4 is grounded ($V_4=0$), one can solve for $V_1=3I/2(h/e^2),V_2=V_6=I(h/e^2),V_3=V_5=I/2(h/e^2)$.
Then the longitudinal resistances $R_{23} = (V_2 - V_3)/I=1/2(h/e^2)$ and $R_{14} = (V_1 - V_4)/I=3/2(h/e^2)$ can be determined, which has exactly been experimentally observed \cite{Roth2009}.

In our numerical calculations, the tight-binding Hamiltonian of the six-terminal QSH effect device can be described by Eq. (\ref{EQ1}).
With the help of the nonequilibrium Green function method \cite{Long2008, Chen2010}, the transmission coefficient $T_{pq}$ can be calculated by
\begin{align}
T_{pq}={\rm Tr}[{\rm \Gamma}_{p} G^{r} {\rm \Gamma}_{q} (G^{r})^{\dagger}],
\end{align}
where ${\rm \Gamma}_{p (q)}(E)=i[\Sigma_{p (q)}^{r}(E)-(\Sigma_{p (q)}^{r}(E))^{\dagger}]$ is the linewidth function for the leads $p (q)$, and $G^{r}(E)=[E-H_{C}-\Sigma_{p=1}^{6}\Sigma_{p}^{r}(E)]^{-1}$ is the retarded Green function corresponding to the Hamiltonian of the central scattering region $H_{C}$.
$\Sigma_{p}^{r}(E)$ is the retarded self-energy due to the coupling to the semi-infinite terminal $p$ which can be calculated numerically \cite{Lee1981}.
Similarly, a small external bias is applied between lead 1 and lead 4, producing a current $I$ flowing along the longitudinal direction.
The transverse terminals 2, 3, 5, and 6 are four voltage probes with their charge currents set to zero, the edge type of the six terminals are
shown in Fig. \ref{fig5}(a). The widths of the terminals 1,4 (2,3,5,6) are $\frac{40a}{\sqrt{3}} (40a)$, and the size of the central scattering area is set to $136a*\frac{40a}{\sqrt{3}}$.
In Fig. \ref{fig5}(b), we plot the longitudinal resistances $R_{14}$ and $R_{23}$ as a function of incident energy $E$.
One can see from Fig. \ref{fig5}(b) that the longitudinal resistances $R_{14}$ and $R_{23}$ exhibit quantized plateaus of $3/2(h/e^2)$ and $1/2(h/e^2)$ near $E=0$, respectively.
It is evident that the two longitudinal resistances satisfy the relationship $R_{14}=3R_{23}=3/2(h/e^2)$ when $E$ is in the bulk gap, which is consistent with our theoretical analysis above.
This result demonstrates the presence of an helical edge state in the bilayer honeycomb lattice nanoflakes.

\section{Conclusions}

We have demonstrated that intrinsic spin-orbit coupling, $C$-type antiferromagnetic order, and staggered potentials can transform the bilayer honeycomb lattice model into a QSH insulator with a pair of helical edge states.
Specifically, the $C$-type antiferromagnetic order breaks the time-reversal symmetry, but opposing staggered potentials both interlayer and intralayer ensure the interlayer $\mathcal{PT}$ symmetry, which is essential for realizing the QSH effect.
The computed bulk band structure, spin Berry curvature, and spin Chern number $\mathcal{C}_s = 1$ confirm the topological nature of the system.
Additionally, We analyze the topological phase diagram of the bilayer system and confirm the stable existence of the QSH effect phase.
The energy band structure of the zigzag boundary and the distribution of spin-momentum-locked helical edge states further confirm the existence of helical edge states.
Finally, we verify these helical edge states through the quantized longitudinal resistances in a six-terminal transport device.
Our results provide strong evidence for realizing the $C$-type antiferromagnetic QSH effect.

\section*{Acknowledgements}
This work was financially supported by the National Natural Science Foundation of China (Grants No. 12074097, No. 12374034 No. 11921005, and No. 12447147),
Natural Science Foundation of Hebei Province (Grant No. A2024205025),
the Innovation Program for Quantum Science and Technology (Grant No. 2021ZD0302403),
the National Key R and D Program of China (Grant No. 2024YFA1409002),
and the China Postdoctoral Science Foundation (Grant No. 2024M760070).

\end{document}